\documentstyle[multicol,aps,epsfig,graphicx,amssymb]{revtex} 

\tighten       

\begin{document}
\draft 

\title{Nonlinear stochastic discrete drift-diffusion 
theory of charge fluctuations and domain relocation times in
semiconductor superlattices}


\author{L. L. Bonilla$^1$\cite{LLB}, O. S\'anchez$^2$\cite{OSR}
and J. Soler$^2$\cite{JSV} }
\address{$^1$Departamento de Matem\'{a}ticas, Escuela
Polit\'ecnica Superior,  Universidad Carlos III
de Madrid,\\ 
Avenida de la Universidad 30, 28911 Legan{\'e}s, Spain\\ 
Also: Unidad Asociada al Instituto de Ciencia de Materiales (CSIC),
28049 Cantoblanco, Spain\\
$^2$ Departamento de Matem\'atica Aplicada,
Facultad de Ciencias, \\
Universidad de Granada, 18071 Granada, SPAIN. }
\date{ \today  }
\maketitle

\begin{abstract}
A stochastic discrete drift-diffusion model is proposed to account for the
effects of shot noise in weakly coupled, highly doped semiconductor
superlattices. Their current-voltage characteristics consist of a number
stable multistable branches corresponding to electric field profiles
displaying two domains separated by a domain wall. If the initial state
corresponds to a voltage on the middle of a stable branch and a sudden
voltage is switched so that the final voltage corresponds to the next
branch, the domains relocate after a certain delay time. Shot noise causes
the distribution of delay times to change from a Gaussian to a first
passage time distribution as the final voltage approaches that of the end
of the first current branch. These results agree qualitatively with
experiments by Rogozia {\it et al} (Phys. Rev. B {\bf 64}, 041308(R) (2001)
).
\end{abstract}
\pacs{73.50.Fq, 73.61.Ey}


\begin{multicols}{2}
\narrowtext 

\section{Introduction}
\label{sec:intro} 
The current-voltage ($I$-$V$) characteristics of highly doped weakly
coupled  semiconductor superlattices (SL) typically exhibits many sharp
branches due to formation of static electric field domains \cite{gra91}.
Two branches are separated by a discontinuity in the current. The electric
field profile associated to a given branch consists of two regions of
constant electric field (domains) separated by a charge accumulation layer
(domain boundary), which is confined to one or several quantum wells. The
location of the domain boundary distinguishes $I$-$V$ branches: as the
voltage increases, the domain boundary is located closer to the injecting
contact and the high field domain increases at the expense of the low
field one \cite{bon95}. Branches exhibit hysteresis cycles due to
coexistence of two or more stable electric field profiles at a given value
of the voltage. Many interesting dynamical phenomena are associated to
these SL: (i) response of the SL to sudden changes in bias (which may force
relocation of electric field domains \cite{kas96,shi97,luo98R,ama01}), and
(ii) self-sustained oscillations of the current provided temperature is
raised or doping is lowered  \cite{kas97,san01}. Motivated by recent
experimental evidence \cite{rog01R,rog01HE}, we shall present in this paper
a stochastic theory of domain relocation in highly doped SL.

In relocation experiments \cite{luo98R,rog01R,rog01}, a doped SL
displaying a multistable $I$-$V$ characteristic is biased (typically) on
the first plateau, say in the middle of a branch. The corresponding field
configuration has two domains separated by a domain wall which is an
accumulation layer. Then the voltage is suddently increased from $V_0$ to
$V_1=V_0 + \Delta V$ and the time evolution of the current is recorded.
Depending on $\Delta V$, the domain wall has to relocate so that a stable
field configuration appropriate to the new voltage is reached
\cite{luo98R}. The outcome has been studied numerically using a discrete
resonant tunneling model with Ohmic boundary conditions \cite{ama01}. For
any $\Delta V<0$ as well as for small positive $\Delta V$, the relocation
of the domain wall always occurs by a direct movement of the charge
monopole forming the domain boundary to its final position. This movement
may be either upstream or downstream the electron flow as needed. However,
for sufficiently large $\Delta V>0$, a charge dipole is injected at the
emitter contact in addition to the existing monopole, because the latter
cannot move upstream beyond one SL period without encountering a stable
field configuration \cite{ama01}. Recent experiments by Rogozia et
al.~\cite{rog01} confirm this theoretical picture. Other experiments have
shown that the relocation time for up jumps ($\Delta V>0$) close to the
discontinuity in the $I$-$V$ characteristic is random and have also
investigated its probability distribution function \cite{rog01R,rog01HE}.
What is causing randomness in the relocation time? In this paper we argue
in favor of shot noise.

 Shot noise occurring during a transport process is due to fluctuations
in the occupation number of states caused by $(i)$ thermal random initial
fluctuations; $(ii)$ the random nature of quantum-mechanical 
transmission/reflection (partition noise). The latter is in turn caused by
the discrete nature of the electric charge.

The rest of the paper is organized as follows. In Section \ref{sec:model},
we derive a stochastic discrete drift-diffusion model (DDD) from the
previously studied deterministic one (see Ref.~\onlinecite{bon00})
considering partition only noise (thermal noise is negligible in the low
temperature limit). The stochastic DDD model has multiplicative white noise
terms obeying Poissonian statistics and it has been solved numerically by
means of a second order scheme proposed by Platen \cite{klo92}. The results
of numerically solving the stochastic model are reported in Section
\ref{sec:results}. Our numerical results agree qualitatively with Rogozia
et al experiments \cite{rog01R}, thereby enforcing the idea that shot noise
is responsible for the observed fluctuations in domain relocation time.
Details on the numerical scheme and comparison to rougher schemes and to
the results of solving the deterministic model with random initial
conditions are contained in the Appendix.

\section{Stochastic Discrete Drift-Diffusion Model}
\label{sec:model}
\begin{figure}
\includegraphics[angle=270,width=8cm]{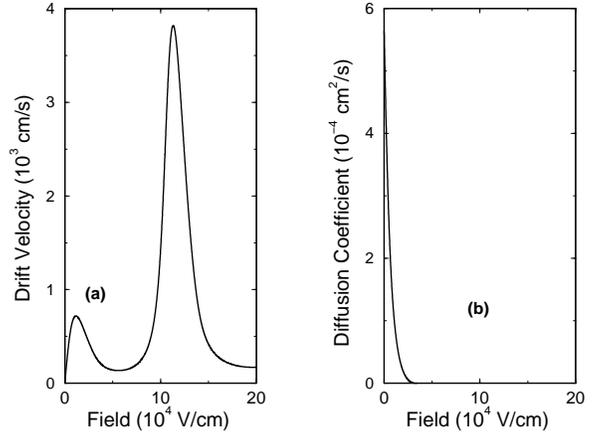}
\caption{\label{fig1} Drift velocity, diffusion coefficient.}
\end{figure}

In weakly coupled SL, typically the scattering times are much shorter than
the escape times from quantum wells. In their turn, the latter are
shorter than typical dielectric relaxation times. This implies that the
dominant mechanism of vertical charge transport is sequential resonant
tunneling and that the tunneling current across barriers can be
considered to be stationary. An appropriate discrete model consists of the
Poisson and charge continuity equations for the two-dimensional electron
density $n_i$ and average electric field $F_i$ at each SL period
\cite{bon95}:
\begin{eqnarray}
F_{i}-F_{i-1} =
\frac{e}{\varepsilon}(n_{i}-N_{D}^{w}) \,\quad  \, i=1, \cdots, N,
\label{Poisson}\\
\frac{d n_{i}}{dt} =  J_{i-1 \to i}-  J_{i \to i+1} \, ,
\quad  \, i=1, \cdots, N .
\label{continuity}
\end{eqnarray}
Here $N_D^w$, $\varepsilon$ and $e J_{i\to i+1}$ are the 2D doping density
at the $i$th well, the average permittivity of the SL and the tunneling
current density across the $i$th barrier, respectively \cite{bon00}. We
can differentiate Eq.\ (\ref{Poisson}) with respect to time and eliminate
$n_i$ by using Eq.\ (\ref{continuity}). The result can be written as a
form of Amp\`ere's law for the balance of current
\begin{eqnarray}
{\varepsilon\over e}\, {dF_{i}\over dt} + J_{i\to i+1} = J(t)\,.
\label{ampere}
\end{eqnarray}
Here $e J(t)$ is the total current density through the SL, equal for all
SL periods, and $\varepsilon\, dF_i/dt$ is the displacement current at the
$i$th SL period. To have a closed system of equations, we need a
constitutive relation linking $e J_{i \to i+1}$ to the unknowns $n_i$ and
$F_i$. For a weakly coupled SL the stationary sequential tunneling current
has been calculated by the Transfer Matrix Hamiltonian method \cite{agu97}
or by the Green function formalism \cite{wac98}. In both cases, for
sufficiently high temperature $J_{i \to i+1}$ may be approximated by a
discrete drift-diffusion law, $J_{i \to i+1}^{(d)} = n_i v(F_i)/l -
D(F_i)\, (n_{i+1}-n_i)/l^2$ \cite{bon00}. The SL period is $l$ and the
electron drift velocity $v(F)$ and the diffusion coefficient $D(F_i)$ have
the forms depicted in Fig. \ref{fig1}.

The DDD model given by Eqs.\ (\ref{Poisson}), (\ref{ampere}) and $J_{i\to
i+1}= J_{i\to i+1}^{(d)}$ has a conceptual difficulty coming from charge
quantization that motivates the introduction of shot noise terms. The
electric charge in each SL period, $e n_i A$ ($A$ is the SL cross section),
should be a multiple of the electron charge $e$. This implies that the true
charge fluctuates about the mean value given by the deterministic DDD
model. To analyze charge fluctuations, we may use the Langevin ideas and
add an appropriate stochastic term to $J_{i\to i+1}^{(d)}$. The SL cross
section $A$ is very large (a circular cross section of diameter 120 $\mu$m
wide as compared to a SL period of $l=13$ nm) and the barrier transmission
coefficient is very small. Then we may use the classic Poissonian shot
noise to model charge fluctuations \cite{bla00}:
\begin{equation}
J_{i \to i+1} = {n_i v^{(f)}(F_i) - n_{i+1} v^{(b)}(F_i)\over l} + J_{i
\to i+1}^{(r)}(t), \label{tun1}
\end{equation}
for $i=1,\cdots N-1$, where  $J_{i\to i+1}^{(r)}$ represents the random
current which satisfies
\begin{equation}
\langle J_{i \to i+1}^{(r)}\rangle = 0, \nonumber
\end{equation}
\begin{eqnarray}
&&\langle J_{i \to i+1}^{(r)}(t) \ J_{j \to j+1}^{(r)}(t')\rangle =
\nonumber\\
&&\delta_{ij} \delta(t-t')\, (Al)^{-1}  [n_i v^{(f)}(F_i) + n_{i+1}
v^{(b)}(F_i)], \label{tun3}
\end{eqnarray}
and $v^{(b)}$, $v^{(f)}$ are defined as follows
\begin{equation}
v^{(b)}(F) = {D(F)\over l}\,,\quad  v^{(f)}(F) = v^{(b)}(F) + v(F),
\label{tun2}
\end{equation}
The logic behind this form of the random tunneling current is as follows.
We consider that uncorrelated electrons are arriving at the $i$th barrier
with a distribution function of time intervals between arrival times that
is Poissonian \cite{bla00}. Then the shot noise spectrum for the current $e
J_{i \to i+1}^{(r)} A$ is given by the average current, $[n_i v^{(f)}(F_i) +
n_{i+1} v^{(b)}(F_i)]\, e^2 A/l$, which in turn yields Eq.\ (\ref{tun3}).
As remarked in Ref.~\cite{bla00}, this procedure assumes low transmission
through barriers and it yields an upper bound for the shot noise amplitude.
In addition, the tunneling current is approximated by a discrete 
drift-diffusion expression whose transport coefficients (drift velocity,
diffusivity, \ldots) will be quantitatively different from those of the
actual sample used in experiments. Given the exponential dependence of
several quantities, relatively small differences in the location of
extrema of the drift velocity, etc.\ may produce substantial differences. 
Thus, the mathematical model provides quantitative differences in the
results but it yields the correct qualitative behavior. 

\begin{figure}
\includegraphics[angle=270,width=8cm]{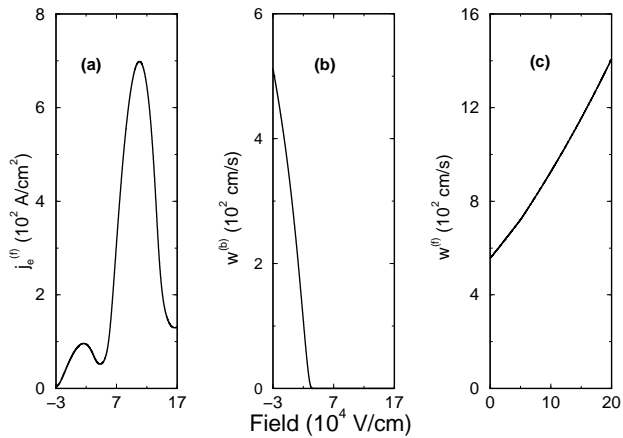}
\caption{\label{fig2} Current--field contact characteristics.}
\end{figure}

The special nature of the emitter and collector layers is considered in
the boundary conditions, given by Eq.\ (\ref{ampere}) with $i=0$ and $i=N$
and different constitutive relations for the tunneling currents
\cite{bon00}:
\begin{eqnarray}
J_{0 \to 1} = j_e^{(f)}(F_0) - {n_{1} w^{(b)}(F_{0})\over l}
+ J_{0 \to 1}^{(r)}\, ,& \label{tun4}\\
J_{N \to N+1}= {n_{N} w^{(f)}(F_{N})\over l} +J^{(r)}_{N \to N+1}\,. &
\label{tun5}
\end{eqnarray}
Here we still have $\langle J_{i \to i+1}^{(r)}\rangle =0$ for $i=0$ and
$i=N$, while the correlations are:
\begin{eqnarray}
 \langle J_{0 \to 1}^{(r)}(t) J_{0\to 1}^{(r)}(t')\rangle =
{j_e^{(f)}(F_0) l+ n_{1} w^{(b)}(F_0)\over Al} \delta(t-t'),
\label{tun6}\\
\langle J_{N \to N+1}^{(r)}(t) J_{N\to N+1}^{(r)}(t')\rangle = {n_{N}
w^{(f)}(F_N)\over A l}\, \delta(t-t'). \label{tun7}
\end{eqnarray}
The emitter current density $e j_{e}^{(f)}$, the emitter backward
velocity $w^{(b)}$ and the collector forward velocity $w^{(f)}$ are
functions of the electric field depicted in Fig.~\ref{fig2} \cite{bon00}.

In addition to the boundary conditions, the Ampere and Poisson equations
should be supplemented with the voltage bias condition,
\begin{equation}
 \sum_{i=1}^{N} F_{i} l = V \, ,
\label{voltcond}
\end{equation}
where $V$ denotes voltage. Eqs.\ (\ref{Poisson}), (\ref{ampere}),
(\ref{tun1}) to (\ref{voltcond}) form a closed system of stochastic
equations for $n_i$, $F_i$ and $J$. They constitute the stochastic DDD
model. To analyze this model, it is convenient to render all equations
dimensionless. Let us denote by $(F_M,v_M)$ the coordinates of the first
positive maximum of the drift velocity $v(F)$. We adopt $F_M$, $N_D^w$,
$v_M$, $v_M l$, $e N_D^w v_M /l$ and $\varepsilon F_M l/(e N_D^w v_M)$ as
units of $F_i$, $n_i$, $v(F)$, $D(F)$, $eJ$ and $t$, respectively. In our
numerical calculations, we have used parameters corresponding to the 9/4 SL
of Ref.~\onlinecite{kas97} at a temperature of 5 K. The three first
subbands have energies of $44$, $180$ and $410$ meV, respectively, and we
assume that the spectral functions of the wells are Lorentzians with
half-widths of 10 meV \cite{bon00}. Then we find $F_M = 11.60$ kV/cm, $N_D^w
= 1.5\times 10^{11}$ cm$^{-2}$, $v_M = 718$ cm/s, $v_M\, l = 9.33\times
10^{-4}$ cm$^2/$s and $eN_D^w v_M/l = 13.27$ A/cm$^2$. For a circular
sample with a diameter of 120 $\mu$m, the units of current and time are
1.501 mA and 1.021 ns, respectively. Then Eqs.\ (\ref{Poisson}),
(\ref{ampere}), (\ref{tun1}) to (\ref{voltcond}) become
\begin{eqnarray}
E_i &-&E_{i-1} = \nu\, (n_i -1)\, , \label{e1}\\
J(t)&=& {dE_i\over dt} + n_i v_i - D_i (n_{i+1}-n_i) \nonumber\\
& & + a \sqrt{n_i (v_i+ D_i)+ D_i n_{i+1}} \, \xi_i(t)\,,
\label{e2}\\
J(t)&=& {dE_{0}\over dt} + J_e(E_0) - W_e(E_0)\, n_1\nonumber\\
& & + a \sqrt{J_e(E_0) + W_e(E_0)\, n_1 } \, \xi_0(t)\,,
\label{e3}\\
J(t)&=& {dE_{N}\over dt}+W_c(E_N)\, n_N \nonumber\\
& & + a \sqrt{W_c(E_N)\, n_N} \,\xi_N(t) \,, \label{e4}\\
\phi &=& {1\over N}\sum_{i=1}^{N} E_i  .  \label{e5}
\end{eqnarray}
Here we have used the same symbols for dimensional and dimensionless
quantities except for the electric field and the coefficient functions in
the boundary conditions. The parameters $\nu = e N_D^w/(\varepsilon F_M)
\approx 1.772$, $\phi = V/(F_M N l)$ and $a= \sqrt{e/(\varepsilon F_M A)}
\approx 3.232\times 10^{-4}$ are the dimensionless doping, the average
electric field (bias) and the noise amplitude respectively. $\xi_i(t)$ is a
zero-mean Gaussian white noise with correlation $\langle \xi_i(t)
\xi_j(t') \rangle =\delta_{ij}\delta(t-t')$ ($\xi_i(t)= \xi_i(t_m)/
\sqrt{\Delta t}$, where the $\xi_i(t_m)$ are independent identically
distributed (i.i.d.) normalized Gaussian
random variables for each discrete time $t_m$ and $\Delta t$ is the
dimensionless time step). The rest of the coefficients in Eqs.\ (\ref{e1})
to (\ref{e4}) are defined by
\begin{eqnarray}
v_i \equiv v(E_i)&=& {v(F_M E_i)\over v_M}\, , \nonumber \\
\vspace{0.2mm}
D_i \equiv D(E_i)&=& {D(F_M E_i)\over V_M l}\, , \nonumber \\
\vspace{0.2mm}
J_e (E_0)&=& { j_e^{(f)}(F_M E_0)l\over N_D^w v_M} \, , \nonumber \\
\vspace{0.2mm}
W_e (E_0)&=& {W^{(b)}(F_M E_0) \over v_M} \, , \nonumber \\
\vspace{0.2mm}
W_c (E_N)&=& {W^{(f)} (F_M E_N) \over v_M}\, .
\end{eqnarray}
The previous system of equations can be further simplified since the
electron densities $n_i$ and the total current density $J(t)$ can be
expressed in terms of the electric field and the bias. Differentiating
Eq.\ (\ref{e2}) with respect to time, and using Eqs.\ (\ref{e3}) and
(\ref{e4}), we obtain an expression for the total current density $J(t)$:
\begin{eqnarray}
{d \phi \over dt} & = &  {1\over N}\,\sum_{i=1}^{N} {dE_i \over dt}
= J - {1\over N}\sum_{i=1}^{N-1} [n_i (v_i +D_i) - n_{i+1} D_i]
\nonumber \\
&- & {n_N W_c (E_N)\over N}- {a\over N}\sum_{i=1}^{N-1} \sqrt{n_i v_i +
(n_i + n_{i+1})D_i }\,\xi_i (t)\nonumber\\
&-& {a\over N} \sqrt{n_N W_c (E_N)}\, \xi_N(t) \, .\nonumber
\end{eqnarray}
Then the total current can be written as
\begin{eqnarray}
& & J =J_1 + {\bf J}_2\cdot {\bf \xi},\label{e6}\\
& & J_1= {d \phi \over dt} + {\sum_{i=1}^{N-1} {n_i (v_i +
D_i) - n_{i+1} D_i\over N}}\nonumber\\
& & \quad\quad\quad\quad\quad\quad\quad\quad\quad\quad\quad\quad\quad\quad
+ {n_N W_c (E_N)\over N} \label{e7}\\  
& & ({\bf J}_2)_0 = 0,\label{e8}\\
& & ({\bf J}_2)_i = {a\,\sqrt{n_i v_i + (n_i + n_{i+1})D_i}\over N}, \quad
1\leq i<N,   \label{e9}\\
& & ({\bf J}_2)_N= {a\,\sqrt{n_N W_c (E_N)}\over N}\,, \label{e10}\\
& & {\bf \xi}= (\xi_0(t),\ldots, \xi_N (t))^T. \label{e11}
\end{eqnarray}
We can now insert these equations in the Amp\`ere equations (\ref{e2}) to
(\ref{e4}) and eliminate $n_i$ by using Eq.\ (\ref{e1}) thereby obtaining
a stochastic differential equation of the following form:
\begin{equation}
{d {\bf E}\over dt} = {\bf H}\left({\bf E},{d\phi\over dt}\right) +
S({\bf E})\cdot {\bf\xi}(t) , \label{sde}
\end{equation}
for the $(N+1)$-dimensional vector electric field ${\bf E} =
(E_0,\dots,E_N)^T$. Here $S$({\bf E}) is a $(N+1)\times (N+1)$ matrix and
{\bf H} is a $(N+1)$-dimensional vector having obvious forms which we do
not write explicitly for the sake of conciseness.

The stochastic differential equation (\ref{sde}) has been numerically
solved by using two different methods: a first order Heun scheme (modified
Euler scheme) and the second order scheme proposed by Platen \cite{klo92}.
The second numerical scheme is rather more costly, but we had to use it to
avoid that numerical errors mask the effects due to charge fluctuations.
Technical details on numerical schemes and a comparison of their
performances are given in Appendix \ref{numerics}. The results of our
simulations are reported in the next Section.
\section{Numerical results}
\label{sec:results}
We have numerically investigated the sample of Ref.~\cite{kas97} that was
used in the relocation experiments \cite{luo98R,rog01R}. It consists of a
$N=40$-period SL with 9-nm wide GaAs wells and 4-nm wide AlAs barriers,
and 2D doping $N_{D}^{w} = 1.5\times 10^{11}$ cm$^{-2}$, at a temperature
$T=5$ K. We have solved numerically the nondimensional equations in the
units and dimensionless parameters introduced in Section \ref{sec:model}.
\begin{figure}
\includegraphics[angle=270,width=8cm]{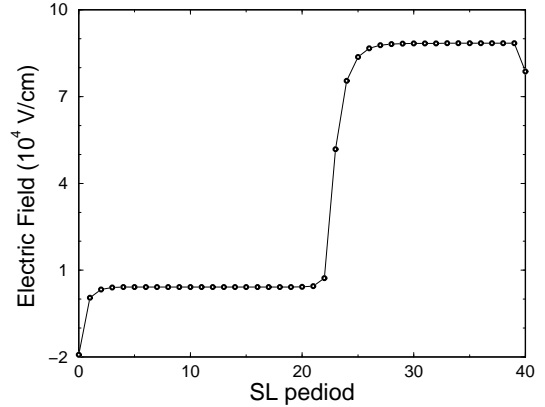}
\caption{\label{fig3} Static electric field
profile at V=$2.1$ V. }
\end{figure}
Figures \ref{fig3} and \ref{fig4} show a typical static electric field
profile (with two coexisting domains) and the first plateau of the time
averaged $I$-$V$ characteristics (obtained by voltage up sweeping). 
To ascertain the influence of charge fluctuations in domain relocation, we
start by setting a stationary field configuration corresponding to a
voltage $V_0= 0.65$ V on the lower branch of Fig.~\ref{fig4}. At time
$t=0$, the voltage increases (in one time step) to its final value $V_f$
on the next $I$-$V$ branch. 
\begin{figure}
\includegraphics[angle=270,width=8cm]{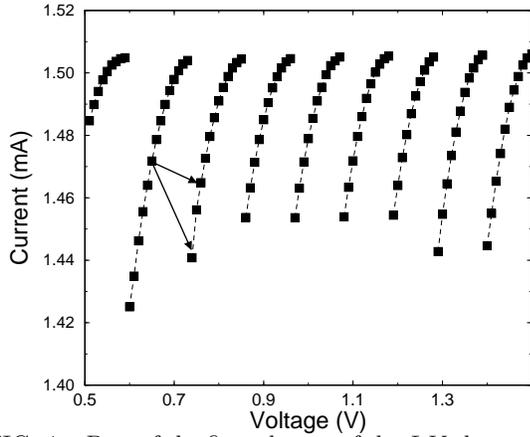}
\caption{\label{fig4} Part of the first plateau
of the $I$-$V$  characteristics. }
\end{figure}
Time traces of the current are depicted in
Figures \ref{fig5} to \ref{fig7}. Notice that the vertical scale has been
augmented sufficiently to see the fluctuations of the current, that are
typically about 0.02 in size. To compare our numerical results to
experimental ones, we need to characterize the domain relocation times and
their distribution function. After a voltage switch, each realization of
the random solution of Eq.\ (\ref{sde}) gives rise to jumps in the mean
current as depicted in Figures \ref{fig5} to \ref{fig7}. We compare the
time trace of the current (time averaged over intervals of five time
dimensionless units) to the value of the current in static $I$-$V$ branches. The first
time $t_0$ that the current time trace differs less than $5\times 10^{-4}$
dimensionless units from its final stationary value, we consider that the
domain relocation has ended. The distribution of time delays $t_0$ taken
over many realizations is then recorded. For a large voltage switch, the
time delay before the current falls from its initial value to its final
level is shorter than for a smaller voltage switch; compare
Figs.~\ref{fig5} and \ref{fig6}. The differences between the time delays
involved in these two cases (about 40 ns) are smaller than those recorded
in experiments \cite{rog01R}.  These differences occur because of
overestimation of the field $F_M$ and the shot noise amplitude by our
theoretical calculations with respect to those of the experimental sample,
as we mentioned before. 
\begin{figure}
\includegraphics[angle=270,width=8cm]{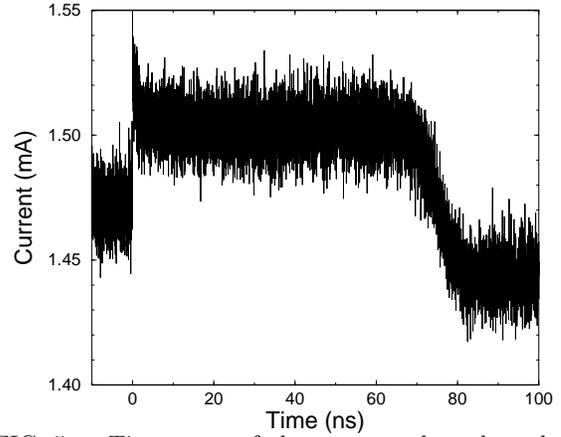}
\caption{\label{fig5} Time trace of the current
when the voltage is  switched from $V_0= 0.65$ V
to $V_f = 0.737$ V. }
\end{figure}
\begin{figure}
\includegraphics[angle=270,width=8cm]{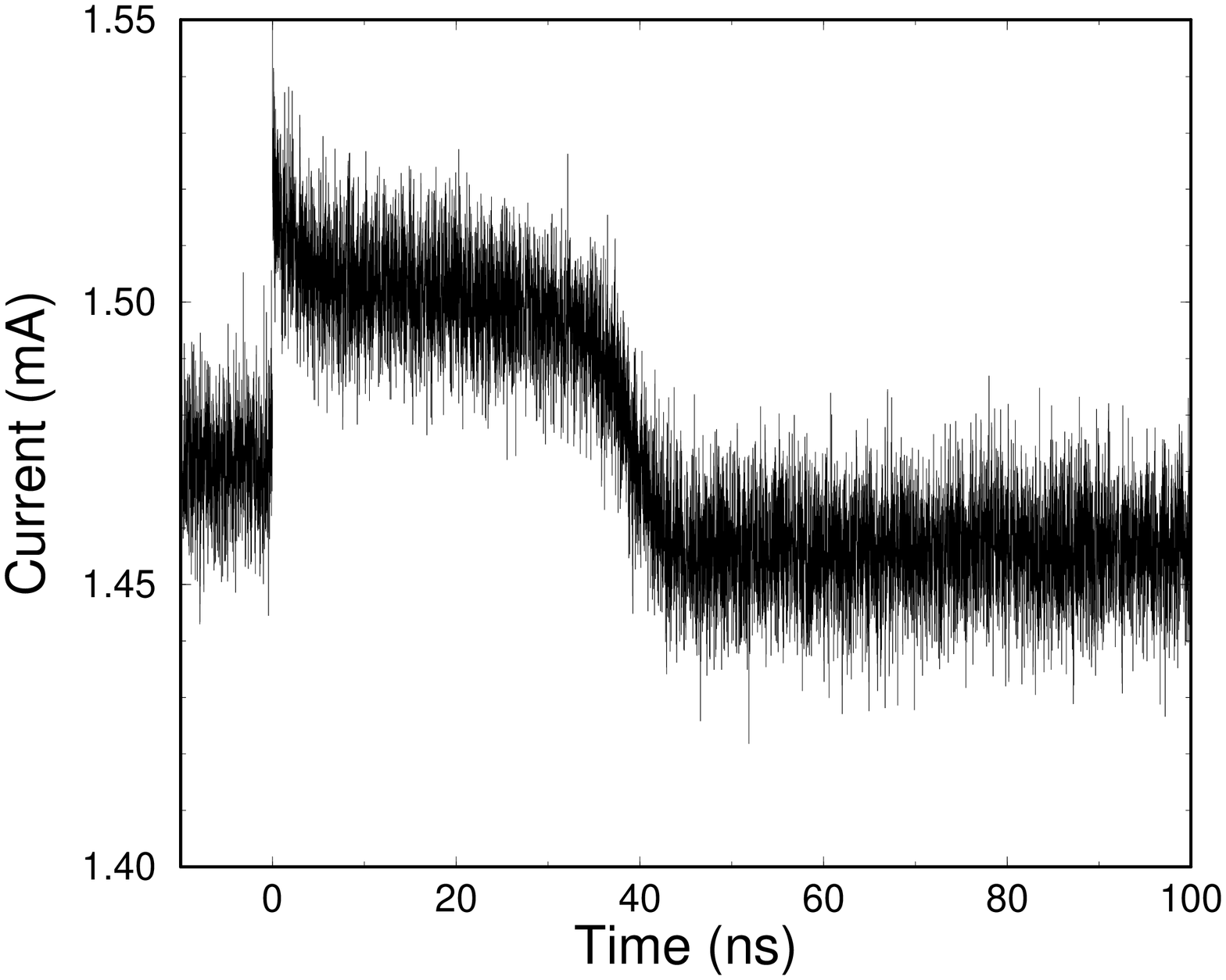}
\caption{\label{fig6} As in Fig.~\protect\ref{fig5}, but with a final
voltage $V_f =0.75$ V.}
\end{figure}
\begin{figure}
\includegraphics[angle=270,width=8cm]{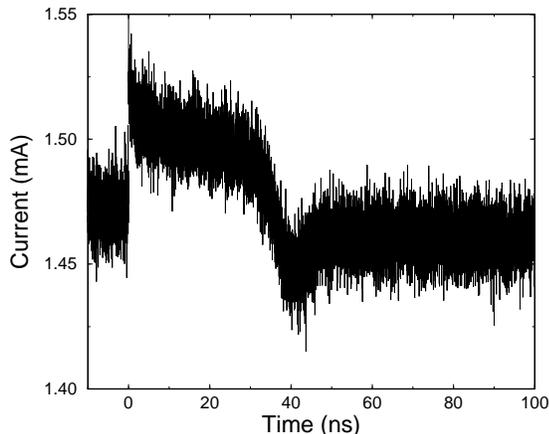}
\caption{\label{fig7} As in
Fig.~\protect\ref{fig5}, but with a final
voltage  $V_f =0.755$ V.}
\end{figure}
In Ref.~\onlinecite{luo98R} it was claimed that the time delay depends
exponentially on the difference between the final value of the stabilized
current, $I$, and the maximum value of the current (or mimimum value in the
case of a down switch) at the initial branch, $I_m$. Then the relocation
time (measured in units of 1.021 ns) depends exponentially on the current
difference $I-I_m$, i.e.,
\begin{equation}
\exp\left({b\, |I-I_m|\over I_M} + c\right)\,.
\end{equation}
We have observed this dependence in our numerical results too. The
dimensionless constants $b$ and $c$ are $b=64.9866$ and $c=1.6717$. $I_M=
1.501$ mA is the unit of current. In Luo et al's experiments
\cite{luo98R}, $I_M = 136 \mu$A (approximately the height of the first
maximum of the current in the inset of Fig.\ 1), $b= 10.74$ (6 times
smaller than the numerically calculated value) and $c=3.34$ (2 times
larger than the numerically calculated value). We thus confirm the
exponential dependence of the relocation time on the current difference
and observe a good qualitative agreement between numerically and
experimentally obtained values.  
\begin{figure}
\includegraphics[angle=270,width=7.7cm]{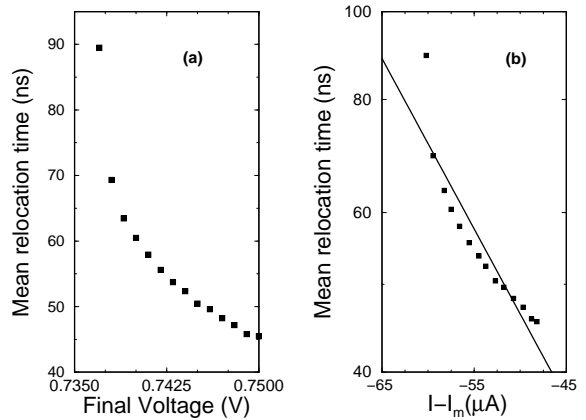}
\caption{\label{fig8} (a) Mean relocation time for different
final voltages. (b) Logarithm of the mean  delay time vs current
difference between final current and the maximum or minimum current
$I_m$ of the initial branch.}
\end{figure}
Fig.~\ref{fig8}(a), shows the mean relocation time obtained in our
simulations as a function of $V_f$. As the final voltage approaches that
corresponding to $I_M$, the relocation time increases. Fig.~\ref{fig8}(b)
depicts the mean relocation time as a function of $(I-I_m)$ on a
semilogarithmic scale for $V_f$ values between $0.737V$ and $0.735V$. The
solid line denotes a linear fit to the data points, that agrees with the
exponential law proposed by Luo et al \cite{luo98R}. These figures are
qualitatively similar to the corresponding ones depicted from experimental
data in Refs.~\onlinecite{rog01R} and \onlinecite{luo98R}. Quantitative
differences are due to the above mentioned discrepancies in $F_M$, the
tunneling current and the shot noise amplitude. Now we focus on the
distribution of switching times. Typically, delay distributions are either
close to symmetric Gaussians or they are asymmetric, depending on how far
$V_f$ is from the limit point of the $I$-$V$ characteristics. We have
fitted our numerical distributions by least squares to either a Gaussian
density:
\begin{equation}
W(t,\tau,\sigma) =
\frac{1}{\sigma \sqrt{2 \pi}}
\exp \left ( -\frac{(t -\tau)^2}{2 \sigma^2} \right ) ,
\end{equation}
or to a first passage time (FPT) distribution
\begin{equation}
W(t,y,\beta) \, dt =
\sqrt{y \frac{2 \beta}{\pi}}
\exp \left( -\frac{\beta y z^2}{2} \right) \, dz \, ,
\end{equation}
where
\begin{equation}
z = \frac{1}{\sqrt{\exp(2 \beta t) -1}} .
\end{equation}
The parameters of these distributions are $\tau$ (mean relocation time) and
$\sigma$ (standard deviation) for the Gaussian and $y$ and $\beta$ for the
FPT distribution. The results of our fitting are depicted in Figures
\ref{fig9} and \ref{fig10}.
\begin{figure}
\includegraphics[angle=270,width=8cm]{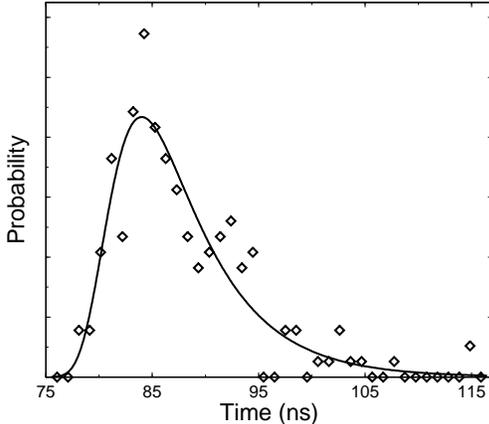}
\caption{\label{fig9} Time delay distribution for $V_f=0.737$ V. Data from
numerical simulations have been fitted to a FPT distribution.}
\end{figure}
\begin{figure}
\includegraphics[angle=270,width=8cm]{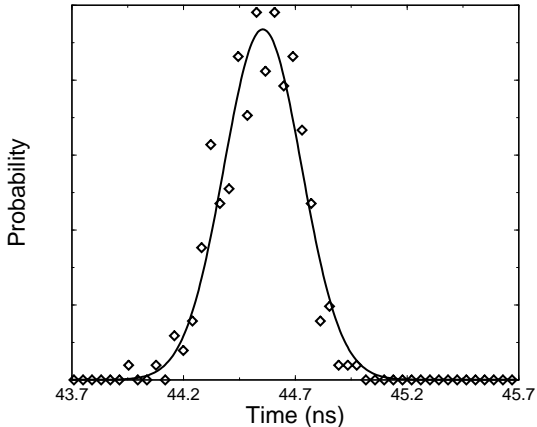}
\caption{\label{fig10} Time delay distribution for $V_f=0.75$ V. Data from
numerical simulations have been fitted to a Gaussian distribution.}
\end{figure}

 These results agree qualitatively with the experimental ones of Rogozia
et al's \cite{rog01R}. As in Ref.~\onlinecite{rog01R}, our Figures
\ref{fig9} and \ref{fig10} show that for values of the voltage far away
from the current jump the time delay distribution changes from an
asymmetric FPT distribution to a very narrow symmetric Gaussian
distribution as $V_f$ departs from the voltage corresponding to the
current jump. These features have a numerical expression in terms of
descriptive statistics like the mean, the standard deviation or the
skewness coefficient  as shown in the Tables of Appendix \ref{numerics}.
The numerically calculated largest and smallest delay times are also
presented.

\section{Conclusions}
\label{sec:concl}
We have studied how the shot noise due to charge quantization affects the
relocation time of electric field domains after a suddent switch of the
voltage. We find that the mean relocation time depends exponentially on
the difference between the value of the current at the final voltage and
the value of the current at the end of the branch corresponding to the
initial voltage. The distribution function of delay times after a voltage
switch changes from Gaussian to a FPT distribution as the final voltage
approaches the limit point of the stationary $I$-$V$ characteristics.
These results are in qualitative agreement with experiments.  

\acknowledgements
We thank Dr.\ Guillermo Carpintero for fruitful discussions during the
early stages of this work.  We acknowledge financial support of the
DGES grants PB98-0142-C04-01 (LLB) and PB98-1281 (OS and JS).

\appendix

\section{Numerical Scheme }
\label{numerics}
\begin{figure}
\includegraphics[angle=270,width=8cm]{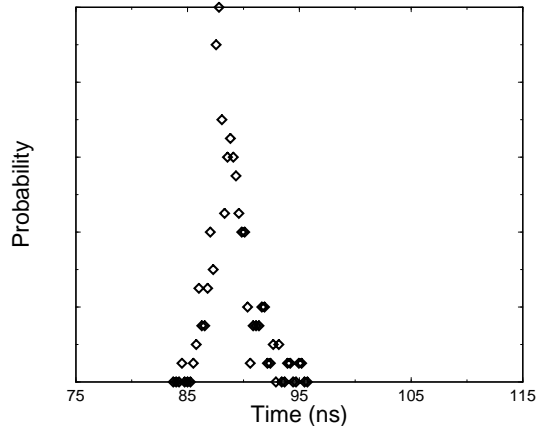}
\caption{\label{fig11}
Time delay distribution for $V_f=0.737$ for different
initial conditions.}
\end{figure}

\begin{figure}
\includegraphics[angle=270,width=8cm]{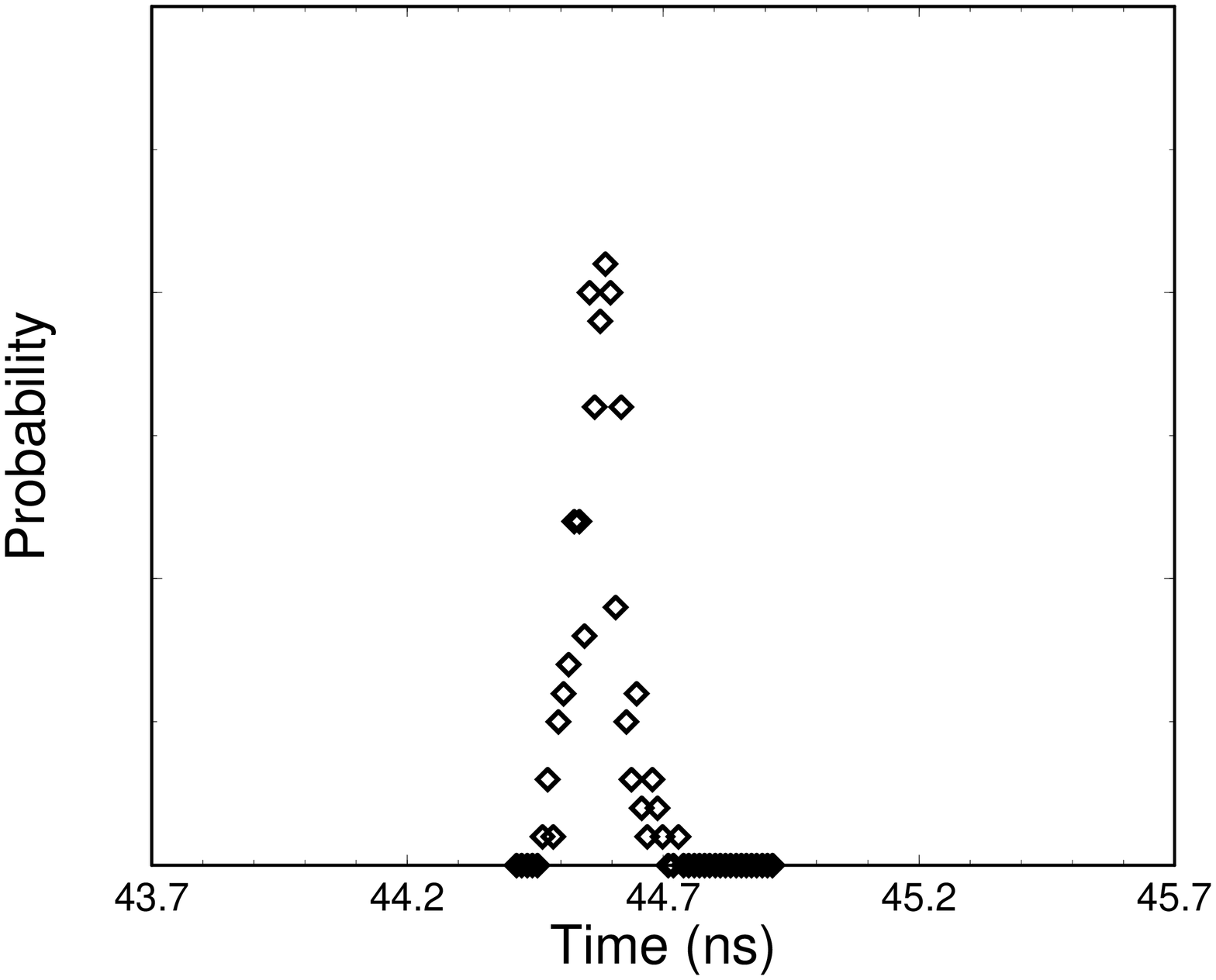}
\caption{\label{fig12}
Time delay distribution for $V_f=0.75$ for different
initial conditions.}
\end{figure}

This Appendix is devoted to explain some technical details of the
simulations. The Platen second order scheme gives the vector field ${\bf
E}^{n+1}$ at discrete time $t+\Delta t$ as the following function of ${\bf
E}^{n}$ at discrete time $t$ \cite{klo92}:
\begin{eqnarray}
{\bf E}^{n+1}= {\bf E}^{n} +
\frac{1}{2} \left({\bf H}\left({\bf \Upsilon},\frac{d \phi}{dt} \right) +
{\bf H}\left({\bf E}^{n},\frac{d \phi}{dt}\right) \right) \Delta t
\nonumber \\
+\frac{1}{4} \sum_{j=1}^{N+1}
\bigg[ \left({\bf S}^j({\bf M}^{j}_{+}) + {\bf S}^j({\bf M}^{j}_{-})+ 2 {\bf
S}^j({\bf E}^{n}) \right) \Delta W^j \nonumber \\
+ \sum_{r=1, r \neq j}^{N+1} 
\left({\bf S}^j({\bf U}_{+}^{r}) + {\bf S}^j({\bf U}^{r}_{-}) - 2 {\bf
S}^j({\bf E}^{n})
\right) \Delta W^j \bigg]   \nonumber \\
+\frac{1}{4} \sum_{j=1}^{N+1}
\bigg[ \left({\bf S}^j({\bf M}^{j}_{+}) - {\bf S}^j({\bf M}^{j}_{-})
\right) \left\{ (\Delta W^j)^2 -\Delta t \right\}
\nonumber \\
+ \sum_{r=1, r \neq j}^{N+1}
\left({\bf S}^j({\bf U}_{+}^{r}) - {\bf S}^j({\bf U}^{r}_{-}) \right)
\left\{ \Delta W^j \Delta W^r + V_{r,j} \right\}
\bigg]\,.
\nonumber
\end{eqnarray}
Here ${\bf S}^j(\cdot)$ is the $j-$th column of $S(\cdot)$, ${\bf
U}_{\pm}= {\bf E}^{n}\pm {\bf S}({\bf E}^n)^j \sqrt{\Delta t}$, and ${\bf
H}$ and $S$ are evaluated at
\begin{eqnarray}
{\bf \Upsilon} = {\bf E}^n + {\bf H} \left({\bf E}^n, \frac{d\phi}{dt}
\right)\Delta t + \sum_{j=1}^{N+1} {\bf S}({\bf E}^n)^j \Delta W^j,
\nonumber \\
{\bf M}^{j}_{\pm} = {\bf E}^{n} +  {\bf H}\left({\bf E}^n, \frac{d\phi}{dt}
\right) \Delta t \pm {\bf S}^j({\bf E}^n) \sqrt{ \Delta t }.    \nonumber
\end{eqnarray}
$\Delta W^j$ are independent gaussian random variables distributed with
zero mean and variance $\Delta t$, whereas the  $V_{j_1,j_2}$ are
independent two point random variables that satisfy
$$
P( V_{j_1,j_2}= \pm \Delta t) = \frac{1}{2}\, ,\quad V_{j_1,j_1} = -\Delta
t \,, V_{j_1,j_2}=- V_{j_2,j_1}\,.
$$
We have used a time step of $\Delta t = 10^{-4}$ (in dimensionless units)
of the same order as the noise amplitude $a$. The values of the random 
variables $V$ and $W$ have been generated through a random number generator
improved by using a seed selector depending on the computer clock and an
algorithm which allows to avoid the sequential correlation usual in this
sort of generators \cite{nrc}. The Platen scheme is second-order weakly
convergent in the following sense. Let $g({\bf E})$ be any sufficiently
smooth scalar function (with $2(\beta +1)$ continuous derivatives provided
$\beta$ is the order of the scheme). Let us fix the time instant at $t$
corresponding to discrete time $n$. Then 
$$
|\langle g({\bf E}^n)\rangle - \langle g({\bf E})\rangle | \leq C
(\Delta t)^{2}\, ,
$$
for any $\Delta t\in (0,\delta_0)$, where $C$ and $\delta_0$ are positive
constants. The Platen numerical scheme is certainly more complicated and
costly than even a stochastic Heun (modified Euler) first order scheme. We
have had to use it to minimize the effects of numerical noise coming from
floating-point arithmetic (even our high-precision 64-bit arithmetic) and
that inherent in interpolating our transport coefficients and contact
functions in the boundary conditions. In fact, in the absence of the noise,
both the Heun and the Platen schemes become the well-known deterministic
Heun (improved Euler) scheme, that is a second-order Runge-Kutta method:
\begin{eqnarray}
{\bf E}^{n+1} = {\bf E}^{n} + \frac{\Delta t}{2}\left[{\bf H}({\bf E}^{n})
+ {\bf H}\big({\bf E}^{n} + {\bf H}({\bf E}^{n})\Delta t\big) \right] \,.  
\nonumber
\end{eqnarray}
However both schemes differ in their treatment of the noise: the
stochastic Heun method is weakly first order whereas the Platen scheme is
second order. The result obtained by using the Platen scheme exhibits less
dispersion than that reached by the Heun method, as shown in Tables
\ref{TableHeun} and \ref{TablePlaten}. An appropriate treatment of the
noise term avoids the presence of artificial numerical effects. The
effects of the numerical perturbations can be illustrated as follows. Let
us use the deterministic Heun scheme with random initial conditions
corresponding to disturbances of the stationary field profile at voltage
$V_0$ and suddenly switch to voltage $V_f$. The domain relocation times
have been measured and they give rise to the distributions of Figures
\ref{fig11} and \ref{fig12}. We have compared the mean, standard deviation
and skewness coefficient
\cite{skewness} of these distributions to those corresponding to the use of
the stochastic Heun and Platen schemes; see Tables \ref{TableHeun},
\ref{TablePlaten} and \ref{TableRIC}. Notice that the mean relocation
times are similar, while the numerical viscosity contributes to scatter
the results. The shot noise does not change the mean values given by the
deterministic model, but the dispersion measured by the standard deviation
increases due to numerical effects (larger in the Heun scheme). The use of
a numerical scheme that reduces these effects is then clearly justified. 
\begin{table}
\caption{Descriptive statistics of the relocation time distributions
obtained with the Heun scheme.}
\label{TableHeun}
\begin{tabular}{lrr}
Heun & $V_{f}= 0.737$ & $V_{f}= 0.75$ \\
\tableline
Lower Limit (ns)& $\ 77.692$ & $43.775$ \\
Upper Limit (ns)& $128.048$ &  $45.633$ \\
Mean (ns)& $\ 87.863$ & $ 44.564$ \\
Standard Deviation (ns)& $\ \ 6.803$ & $\ 0.299$ \\
Skewness coeff. & $\ \ 1.840$ & $\ 0.131$
\end{tabular}
\end{table}
\begin{table}
\caption{\label{TablePlaten} Descriptive statistics of the relocation 
time distributions obtained with the Platen scheme. }
\begin{tabular}{lrr}
Platen & $V_{f}= 0.737$ & $V_{f}= 0.75$ \\
\tableline
Lower Limit (ns) & $\ 77.827$ & $43.942$ \\
Upper Limit (ns)& $115.025$ &  $44.960$ \\
Mean (ns)& $\ 87.635$ & $ 44.541$ \\
Standard Deviation (ns) & $\ \ 6.339$ & $\ 0.167$ \\
Skewness coeff. & $\ \ 1.4237$ & $-0.2791$
\end{tabular}
\end{table}

\begin{table}
\caption{\label{TableRIC}Descriptive statistics of the relocation time
distributions obtained with pertubed initial conditions.}
\begin{tabular}{lrr}
& $V_{f}= 0.737$ & $V_{f}= 0.75$ \\
\tableline
Mean (ns) & $88.916$ & $ 44.579$ \\
Standard Deviation (ns) & $\ 1.773$ & $\ 0.045$ \\
Skewness coeff. & $\ 0.912$ & $\ 0.255$
\end{tabular}
\end{table}

\end{multicols}

\begin{references}{}
\bibitem{skewness}
The skewness or Fisher coefficient measures the asymmetry of a distribution
and is calculated as the distribution third central moment divided by the
third power of its standard deviation.
\end{references}
\end{document}